\documentclass[english,
longauth]{aa}
\usepackage{epsf,amsfonts,amssymb,graphicx,fancyheadings,caption}
\usepackage{natbib,longtable}
\usepackage[latin1]{inputenc}
\usepackage[active]{srcltx}
\usepackage{babel}
\bibpunct{(}{)}{;}{a}{}{,}

\begin{document}

\title{The beat Cepheids in the Magellanic Clouds: an analysis 
from the EROS-2 database
\thanks{Based on observations made by the EROS-2 collaboration with the MARLY 
telescope at the European Southern Observatory, La Silla, Chile.}}

\author{
J.B.~Marquette\inst{1},
J.P.~Beaulieu\inst{1},
J.R.~Buchler\inst{2},
R.~Szab{\'o}\inst{3},
P.~Tisserand\inst{4,5},
S.~Belghith\inst{1},
P.~Fouqu\'e\inst{6},
\'E.~Lesquoy\inst{5,1},
A.~Milsztajn\inst{5}\fnmsep\thanks{deceased},
A.~Schwarzenberg-Czerny\inst{7,8},
C.~Afonso\inst{3,9},
J.N.~Albert\inst{10},
J.~Andersen\inst{11},
R.~Ansari\inst{10},
\'E.~Aubourg\inst{5},
P.~Bareyre\inst{5},
X.~Charlot\inst{5},
C.~Coutures\inst{5,1},
R.~Ferlet\inst{1},
J.F.~Glicenstein\inst{5},
B.~Goldman\inst{5,9},
A.~Gould\inst{12},
D.~Graff\inst{12,13},
M.~Gros\inst{5},
J.~Ha\"{\i}ssinski\inst{10},
C.~Hamadache\inst{5},
J.~de Kat\inst{5},
L.~Le Guillou\inst{5,14},
C.~Loup\inst{1,15},
C.~Magneville\inst{5},
\'E.~Maurice\inst{16},
A.~Maury\inst{17,18},
M.~Moniez\inst{10},
N.~Palanque-Delabrouille\inst{5},
O.~Perdereau\inst{10},
Y.R.~Rahal\inst{10},
J.~Rich\inst{5},
M.~Spiro\inst{5},
A.~Vidal-Madjar\inst{1}
}
\institute{
%1
Institut d'Astrophysique de Paris, UMR7095 CNRS, Universit\'e Pierre \& Marie Curie, 98~bis Boulevard Arago, 75014 Paris, France
\and
%2
Physics Department, University of Florida, Gainesville, FL 32611-8440, USA
\and
%3
Konkoly Observatory, Budapest, P.O. Box 67, H-1525, Hungary
\and
%4
Research School of Astronomy \& Astrophysics, Mount Stromlo Observatory, Cotter Road, Weston ACT 2611, Australia
\and
%5
CEA, DSM, DAPNIA, Centre d'\'Etudes de Saclay, 91191 Gif-sur-Yvette Cedex, France
\and
%6
Observatoire Midi-Pyr\'en\'ees, Laboratoire d'Astrophysique (UMR 5572), 14 av. E. Belin, 31400 Toulouse, France
\and
%7
Centrum Astronomiczne im. M. Kopernika, Bartycka 18, 00-716 Warszawa, Poland
\and
%8
Obserwatorium Astronomiczne, Uniwersytet A. Mickiewicza, Sloneczna 36, 60-286 Poznan, Poland
\and
%9
Max-Planck-Institut f\"ur Astronomie, K\"onigstuhl 17, D-69117 Heidelberg, Germany
\and
%10
Laboratoire de l'Acc\'el\'erateur Lin\'eaire, IN2P3 CNRS, Universit\'e de Paris-Sud, 91405 Orsay Cedex, France
\and
%11
The Niels Bohr Institute, Copenhagen University, Juliane Maries Vej 30, DK2100 Copenhagen, Denmark
\and
%12
Department of Astronomy, Ohio State University, Columbus, OH 43210, U.S.A.
\and
%13
Division of Medical Imaging Physics, Johns Hopkins University, Baltimore, MD 21287-0859, USA
\and
%14
LPNHE, CNRS-IN2P3 and Universit\'es Paris 6 \& Paris 7, 4 place Jussieu, 75252 Paris Cedex 05, France
\and
%15
Observatoire Astronomique de Strasbourg, UMR 7550, 11 rue de l'Universit\'e, 67000 Strasbourg, France
\and
%16
Observatoire de Marseille, 2 Place Le Verrier, 13248 Marseille Cedex 04, France
\and
%17
European Southern Observatory (ESO), Casilla 19001, Santiago 19, Chile
\and
%18
San Pedro de Atacama Celestial Exploration, Casilla 21, San Pedro de Atacama, Chile
}

\date{Received ; Accepted}
\abstract
{A number of microlensing dark-matter surveys have produced tens 
of millions of light curves of individual background stars. These data provide an unprecedented opportunity for systematic studies 
of whole classes of variable stars and their host galaxies.}
{We aim to use the EROS-2 survey of the Magellanic Clouds to 
detect and study the population of beat Cepheids (BCs) in both 
Clouds. BCs pulsating simultaneously in the first overtone and 
fundamental modes (FO/F) or in the second and first overtone 
modes (SO/FO) are of particular interest.}
{Using special software designed to search for periodic variables, we have scanned the EROS-2 data base for variables in the typical period range of Cepheids. Metallicities of FO/F objects were then calculated from linear nonadiabatic convective stellar models.}
{We identify 74 FO/F BCs in the LMC and 41 in the SMC, and 173 and 129 SO/FO pulsators in the LMC and SMC, respectively; 185 of these stars are new discoveries. For nearly all the FO/F objects we determine minimum, mean, and maximum values of the metallicity.}
{The EROS data have expanded the samples of known BCs in the LMC by 31\%, in the SMC by 110\%. The FO/F objects provide independent measures of metallicities in these galaxies. The mean value of metallicity is 0.0045 in the LMC and 0.0018 in the SMC.}


\authorrunning{J.B. Marquette et al.}
\titlerunning{An analysis of beat Cepheids in the Magellanic Clouds}

\maketitle

\section{Introduction}
\label{sec:Intro}

Over the last decade, large microlensing surveys have been conducted to search for planetary- to stellar-mass baryonic dark matter in the Milky Way. 
While, interestingly, the original search was in fact negative \citep{2007A&A...469..387T}, the resulting tens of millions of homogeneous, 
well-sampled light curves of the background stars have stimulated many new programmes on variable stars. In particular, specific stellar populations like Cepheids can now be studied far more thoroughly, particularly in the Magellanic Clouds. Among these objects, the subsample of double-mode or beat Cepheids (hereafter BCs) is of particular importance in gaining information on the structure and evolution of massive stars. 

According to the General Catalogue of Variable Stars \citep{2004yCat.2250....0S}, the Galactic sample of known BCs is rather small: up to now, only 16 objects of this type have been identified. 
In the Large (LMC) and Small (SMC) Magellanic Clouds the situation is more favourable, thanks in particular to the work of the 
EROS-1 \citep{1995A&A...303..137B,1997A&A...321L...5B},
MACHO \citep{1995AJ....109.1653A,1999ApJ...511..185A}, and OGLE
\citep{1999AcA....49....1U,2000AcA....50..451S} groups. EROS-1 detected one BC in the LMC and 11 in the SMC, while the MACHO group found 45 BCs in the LMC. The OGLE collaboration reported the discovery of 93 BCs in the SMC and 76 in the LMC, very recently extended by the OGLE-III data to a region of 40 $\deg^2$ of the LMC \citep{2008arXiv0808.2210S}. This thorough analysis yielded 266 BCs of different types discussed below. The present study will report results on an even larger region of 88 $\deg^2$ in the LMC and 10 $\deg^2$ in the SMC. 

A BC pulsates either in the first overtone and fundamental modes (FO/F), or in the second and first overtone modes (SO/FO)\footnote{Two cases are known of pulsations in the first and third overtones \citep{2008arXiv0808.2210S}.}. The studies listed above clearly established that the period ratio (higher to lower mode) of the FO/F pulsators is around 0.72, while that of SO/FO objects is near 0.80. For the FO/F pulsators it is well known that the period ratio depends on the metallicity $Z$, as seen e.g. in the Petersen diagram \citep{1973A&A....27...89P} of the period ratio vs. the period of the lower mode.  From OGLE photometry and stellar atmosphere models, \citet{2008arXiv0802.4166K} showed that, in both Clouds, the average metallicity of FO Cepheids is lower than those pulsating in the fundamental mode.

The chemical enrichment history of the Magellanic Clouds has been recently reexamined by \citet{2008AJ....136.1039C, 2008AJ....135..836C}, who used the infrared CaII triplet as a metallicity indicator for red giant branch stars \citep{2007AJ....134.1298C}. \citet{2006A&A...456..451L} used planetary nebulae to trace the heavy elements in the Magellanic Clouds. They concluded that only Ar is a suitable tracer of the metallicity of the progenitor stars, instead of an average of several different elements. \citet{2007A&A...472..101I} also studied the chemical evolution of the SMC from planetary nebulae and concluded that the star formation histories of the LMC and SMC are different from each other and in many respects still controversial.

The situation is complicated by the fact that metallicity measurements in the Magellanic Clouds yield many different values, depending both on the location and age of the stars studied. Thus, a single firm value of the metallicity in these galaxies cannot be assigned. Abundances of $\alpha$-, iron-peak and 
$s$-process elements in individual stars are now being measured with 
high-resolution spectrometers, for example by \citet{2008A&A...480..379P}, but it remains important to contribute to these efforts by adding a new and independent way to measure the metallicity, as we do below.

In a similar study, \citet{2006ApJ...653L.101B} deduced a metallicity gradient in M33, only from the pulsational properties of 5 newly 
discovered BCs in M33 and with the help of stellar pulsation theory and 
mass-luminosity relations derived from evolutionary tracks. The result was
in good agreement with the standard spectroscopic metallicity gradients
as determined from H II regions, early-B supergiant stars, and planetary
nebulae. BCs are thus a new and independent probe of galactic metallicity. In this paper we present the results of a systematic search for and analysis of BCs in the Magellanic Clouds from the EROS-2 database. Numerous previously unknown objects 
are detected, and metallicities are derived from the properties of FO/F BCs in both Clouds.

\section{Observational setup and BC search}
\label{sec:Observ}

The EROS-2 experiment was conducted between July 1996 and March 2003 using 
the dedicated MARLY telescope (1m Ritchey-Chr\'etien, f/5.14) at ESO,
La Silla. A dichroic beam-splitter allowed simultaneous imaging in two
broad non-standard passbands, $B_{\rm E}$ (4200-7200
\AA{}, ``blue'' channel) and $R_{\rm E}$ (6200-9200 \AA{}, ``red''
channel). Each camera contained a mosaic of eight 2K $\times$ 2K LORAL CCDs
with a pixel size of 0.6\arcsec and a field of view of 0.7\degr (right
ascension) $\times$ 1.4\degr (declination).

The light curves of individual stars were constructed from fixed positions on
templates using special-purpose software for photometry of EROS-2 images called PEIDA
\citep{1996VA.....40..519A}. This resulted in a database of tens of millions of light curves, which will be publicly available in the future from the CDS through a dedicated server. For a typical object, 800 to 1000 epochs of observation are available.

Each light curve is named according to the rules given by \citet{2002A&A...389..149D}.  As an example, the star J053929-703819 with 
J2000 equatorial coordinates of 05:39:29.15, -70:38:19.2 is also referred 
to by the internal EROS-2 number lm0591l5619 (the 5619$^{th}$ star in the field lm059, CCD 1, quarter l). In the following, we refer to the objects 
by their internal number. 

A systematic search for periodic objects using the $B_{\rm E}$ observations was performed among stars detected as potentially variable by
an automatic time series analysis pipeline, based on software developed by
\citet{1997A&A...321L...5B} and \citet{2003ASPC..292..383S}, and on an
Analysis of Variance (AoV) method. For each periodic pulsator we constructed a Fourier decomposition of the light curve, $i.e.$ the signal was 
represented by 
$R(t) = R_{0} + \sum_{i = 1}^{i = 5} R_{i}~ \mathrm{cos}(\frac{2\pi}
{P}i(t-t_0) + \Phi_{i})$ \citep{1981ApJ...248..291S}. 
The quantities $R_{k1} = R_{k}/R_{1}$ and $\Phi_{k1} =
\Phi_{k} - k~\Phi_{1}$ ($1 < k \leq 3$) were defined as usual. 

The maximum of the statistic $\Theta_{AoV}$ as defined by \citet{2003ASPC..292..383S} was used as a quality parameter for the periodograms, which were stored when $\Theta_{AoV} \gtrsim 20$. In this way, a total of about 80\,000 variables was found in both Clouds. The Fourier model was then subtracted from the observational data before a second step 
of period search by the same
pipeline. Double-mode objects thus emerged quite clearly from the resulting
periodograms. Potential BCs were identified by having period ratios near 
0.73 and 0.80 for FO/F and SO/FO pulsators, respectively. 

To construct the final samples, the light curves of all candidates (folded according to the first mode of pulsation) were inspected visually, using the versatile tool TOPCAT developed by the Starlink
consortium\footnote{Available at 
\texttt{http://www.star.bris.ac.uk/$\sim$mbt/topcat/}} and
customized to display SuperMongo plots. On a typical BC light curve, folded according to either the first or second period, the beat phenomenon is easily recognised because the light curve is broader than that of a single mode Cepheid. This was our ultimate criterion of selection for objects with the lowest values of $\Theta_{AoV}$. 

\begin{figure*}
  \resizebox{\hsize}{!}{\includegraphics{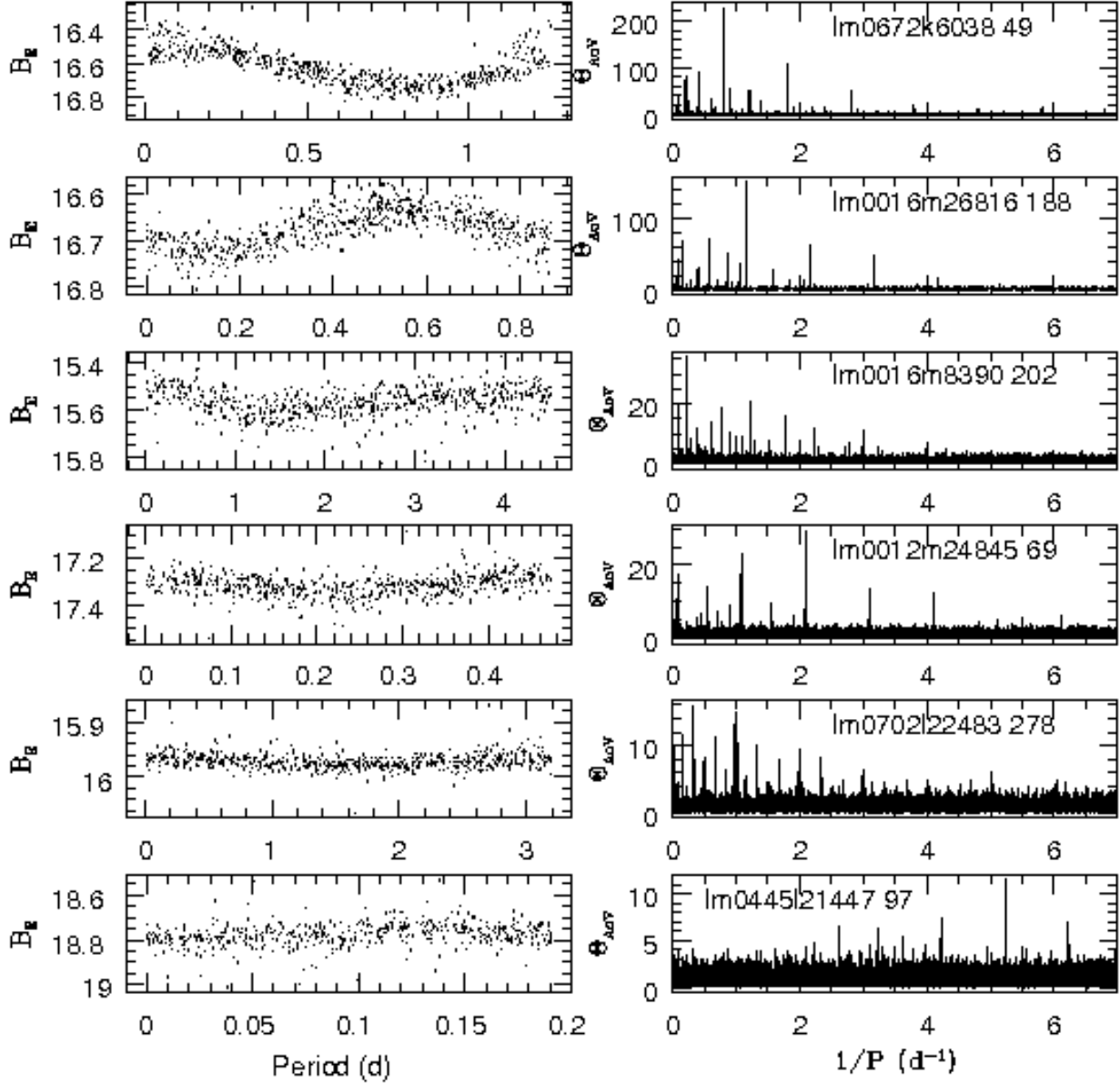}}
\caption{Folded light curves and corresponding periodograms of 6 typical LMC BCs, alternately FO/F and SO/FO objects, from the most significant (top) to non-significant secondary modes (bottom). Left: $B_{\rm E}$ magnitude vs. period (days); right: $\Theta_{AoV}$ for the second pulsation vs. frequency (days$^{-1}$). The name of each object is given in the right-hand panels, together with the maximum $\Theta_{AoV}$ for the first pulsation mode. See text for details.}
\label{fig:sample} 
\end{figure*}

Figure \ref{fig:sample} shows six typical LMC objects, alternately of FO/F and SO/FO type. The two first are firm detections with high values of $\Theta_{AoV}$ for the second pulsation mode, the two in the middle are detected more marginally, and the two last have been rejected as having too low values of $\Theta_{AoV}$. The left-hand panels show the folded light curves (blue magnitude vs. period of the second mode of pulsation, in days), after subtraction of the Fourier model of the first mode. The right-hand panels show the corresponding periodograms ($\Theta_{AoV}$ vs. frequency of the second mode period). In each of these periodograms the highest peak corresponds to the most probable frequency.

\section{General analysis and determination of the metallicities}
\label{sec:metals}

The visual inspection yielded 74 FO/F pulsators in the LMC and 41 in the
SMC as well as 173 SO/FO stars in the LMC and 129 in the SMC. The pertinent data for these stars are given in separate tables for the FO/F and SO/FO pulsators in each galaxy (LMC and SMC) which are only available in electronic form at the CDS via anonymous ftp to cdsarc.u-strasbg.fr (130.79.128.5) or via http://cdsweb.u-strasbg.fr/cgi-bin/qcat?J/A+A/. Table \ref{tab:samples} summarizes
their contents.

\begin{table}
\caption{Contents of the data tables for the new BCs in the LMC and SMC.}
\label{tab:samples} 
\begin{tabular}{ll}
 \hline \hline 
 Column Name & Comments \\
 \hline 
 EROS-2 ID & Object identifier$^{a}$ \\
 RA & J2000 right ascension \\
 DEC & J2000 declination \\
 $P_{0}$ & Period (days) of the $1^{st}$ mode of pulsation \\
 $P_{1}$ & Period (days) of the $2^{nd}$ mode of pulsation \\
 $P_{10}$ & Period ratio $P_{1}/P_{0}$\\
 $Z_{\mathrm max}$ & Maximum metallicity$^{b}$ \\
 $Z_{\mathrm min}$ & Minimum metallicity$^{b}$ \\
 $Z_{\mathrm mean}$ & Mean metallicity$^{b}$ \\
 $R_{0}$ & Fourier coefficient in $B_{\rm E}$  \\
 $R_{1}^{1}$ & \textit{idem}$^{c}$ \\
 $\Phi_{1}^{1}$ & \textit{idem}$^{c}$ \\
 $R_{21}^{1}$ & \textit{idem}$^{c}$ \\
 $\Phi_{21}^{1}$ & \textit{idem}$^{c}$ \\
 $R_{31}^{1}$ & \textit{idem}$^{c}$ \\
 $\Phi_{31}^{1}$ & \textit{idem}$^{c}$ \\
 $R_{1}^{2}$ & \textit{idem}$^{c}$ \\
 $R_{21}^{2}$ & \textit{idem}$^{c}$ \\
 $\Phi_{21}^{2}$ & \textit{idem}$^{c}$ \\
 $R_{31}^{2}$ & \textit{idem}$^{c}$ \\
 $\Phi_{31}^{2}$ & \textit{idem}$^{c}$ \\
 \hline 
\end{tabular} \\
$^{a}$as defined by \citet{2002A&A...389..149D} \\
$^{b}$for FO/F pulsators only \\
$^{c}$superscript $i$ refers to the $i^{th}$ pulsation mode
\end{table} 

\begin{figure*}
  \resizebox{\hsize}{!}{\includegraphics{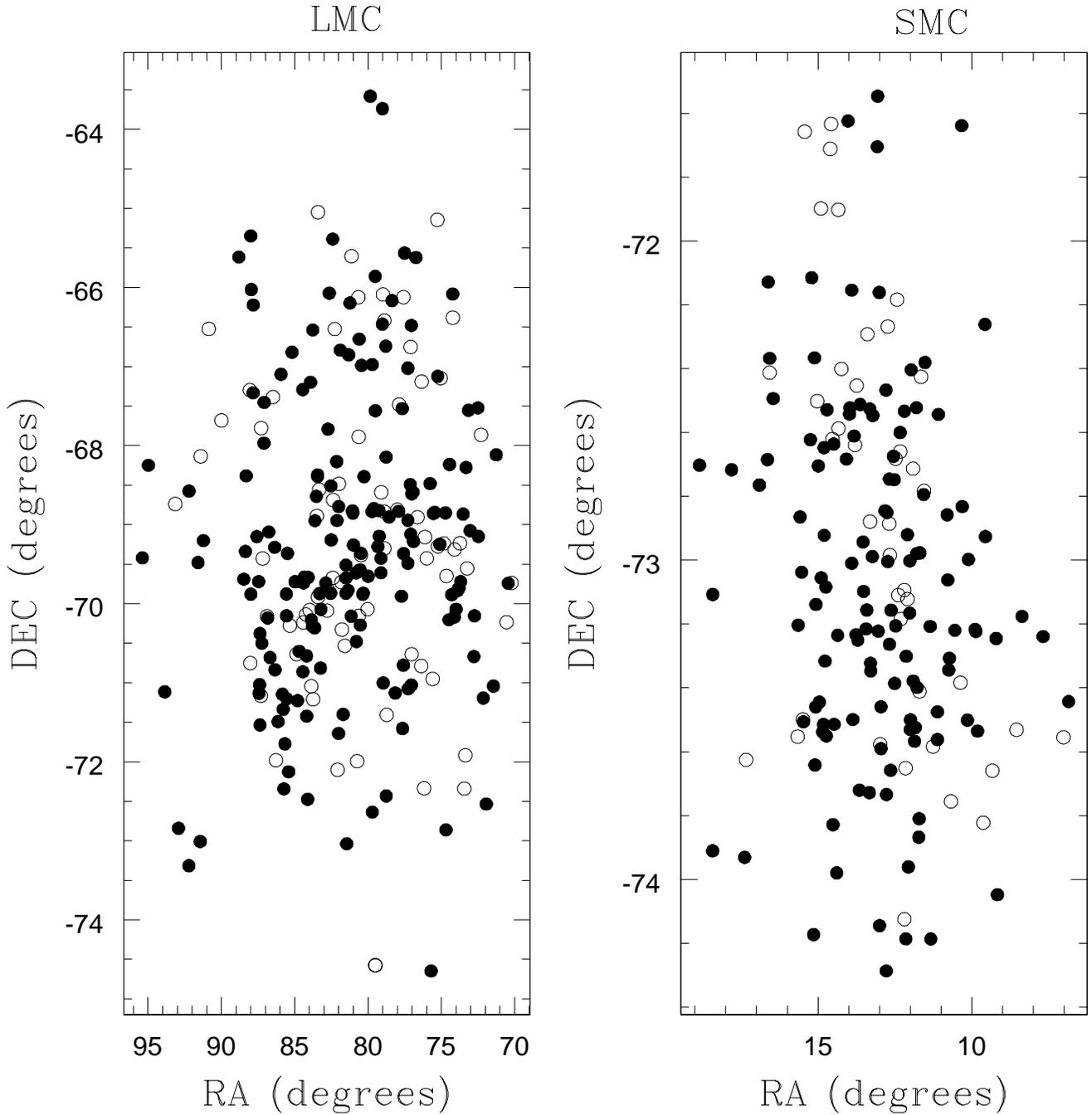}}
\caption{Spatial distribution of BCs in the Magellanic Clouds. Open circles: FO/F objects; filled circles: SO/FO objects.}
\label{fig:spatdist} 
\end{figure*}

Figure \ref{fig:spatdist} displays the spatial distribution of our BCs in both Clouds. The bar of the LMC is clearly visible, although FO/F and SO/FO objects are widely distributed in both the LMC and SMC. We note a surdensity of Cepheids in the northern part of the LMC ($75° < \alpha < 90°$, $-68° < \delta < -66°$), corresponding to a region of star formation. We have checked that BCs have a spatial distribution similar to that of single-mode Cepheids by performing Kolmogorov-Smirnov (KS) tests using the \texttt{stats} package of R, a language and environment for statistical computing \citep{2008sf}. We computed the mean coordinates of the barycentre for the subsamples of F and FO objects, obtained to first approximation from a visual cut in the period-luminosity relation, and for the BCs. We then calculated the distance to the barycentre for each object and made the one-sided KS tests reported in Table \ref{tab:KS} on the resulting distance distributions. 

\begin{table*}
\caption{Results of the one-sided Kolmogorov-Smirnov tests performed on the spatial distributions of two different sets of Cepheids, noted (a) and (b).  See text for further detail.}
\label{tab:KS}
\begin{tabular}{|c|c|c|c|c|}
\hline 
 & Set (a) & Set (b) & $D^b$ & p-value$^c$\tabularnewline
\hline
\hline 
LMC & F (1772)$^a$ & FO (907) & 0.0455 & 0.1660\tabularnewline
 & F + FO (2679) & FO/F + SO/FO (247) & 0.0994 & 0.0245\tabularnewline
\hline 
SMC  & F (1256) & FO (1281) & 0.0568 & 0.0335\tabularnewline
 & F + FO (2537) & FO/F + SO/FO (161) & 0.1322 & 0.0107\tabularnewline
\hline
\end{tabular}\\
$^a$Numbers in parentheses are the sizes of the populations \\
$^bD$ defines the maximum distance between the distance distributions of (a) and (b) \\
$^c$The p-values are the probability that the distributions are different
\end{table*} 

We first verified that the distributions of F and FO Cepheids are almost identical (first and third lines in Table \ref{tab:KS}). Note however the relatively high p-value (i.e. the probability to reject the null hypothesis) of this KS test in the LMC. The low p-values in the second and fourth lines in Table \ref{tab:KS} show that BCs and single mode Cepheids are distributed similarly. The $D$ statistic represents the maximum distance between the distance distributions of populations (a) and (b).

We have compared our findings with the most recent studies from the OGLE group. Table \ref{tab:compare} summarizes the numbers of stars which are in common between EROS-2 and OGLE, and those which were found only by EROS-2 or OGLE, respectively. For both the FO/F or SO/FO samples, the first column list the number of objects detected by both surveys, while the next two columns give the number of stars that are unique to each. OGLE data for the LMC are from \citet{2008arXiv0808.2210S}, for the SMC from \citet{1999AcA....49....1U}. Although a detailed analysis of this comparison is beyond the scope of the present paper, it appears that a large fraction of BCs are detected by both experiments, especially in the LMC for which an extended OGLE-III sample is available \citep{2008arXiv0808.2210S}. The present work has, however, significantly increased the known samples of BCs, by some 31\% in the LMC and 110\% in the SMC. 

There can be many reasons why some objects appear in the EROS-2 data and not in OGLE and vice versa. E.g., the spatial coverage is not the same, and the EROS-2 photometry is PSF-based while OGLE uses a differential image analysis method. In addition, the periodicity search algorithms are different and it should be useful to study whether they are sensitive to different numerical effects or not. In the LMC, 66 (22 FO/F and 42 SO/FO) of the 82 objects that are detected by EROS-2 only are located in regions out of the OGLE-III fields. In the SMC, 69 (17 FO/F and 52 SO/FO) of the 81 objects identified only by EROS-2 are external to the OGLE-II fields. Inversely, on the 127 BCs found by OGLE only, we estimated than 40\% are located either outside or on the edge of the EROS-2 CCDs, which did not allow a proper photometric follow-up. For the 60\% remaining, a mixed between EROS-2 lower photometric accuracy and difference in algorithm sensitivity is thought to be the answer. Clearly, this question should be examined in more details in the future.

\begin{table*}
\caption{Comparison of the EROS-2 (this work) and OGLE BC samples in the LMC and SMC. }
\label{tab:compare} 
 \begin{tabular}{c|c|c|c|c|c|c}
\cline{2-4} \cline{5-7} 
 & \multicolumn{3}{c|}{FO/F} & \multicolumn{3}{c}{SO/FO}\tabularnewline
\cline{2-4} \cline{5-7} 
 & EROS-2 \& OGLE & EROS-2 & OGLE & EROS-2 \& OGLE & EROS-2 & OGLE \tabularnewline
\hline 
LMC & 45 & 29 & 16 & 120 & 53 & 83\tabularnewline
SMC & 18 & 23 & 5 & 48 & 81 & 23\tabularnewline
\hline
\end{tabular}
\end{table*} 

\begin{figure}
 \resizebox{\hsize}{!}{\includegraphics{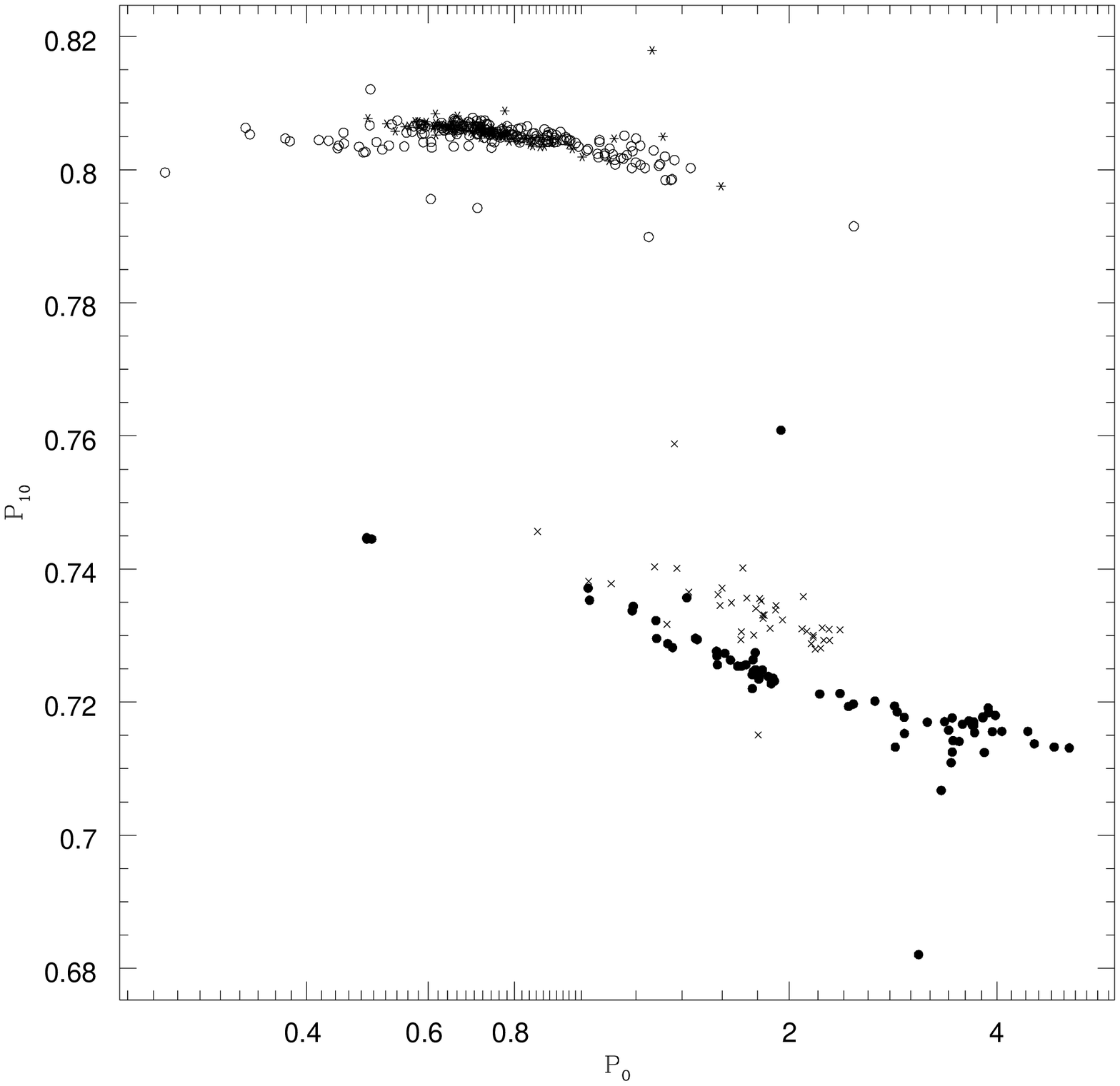}}
\caption{Petersen diagram of FO/F EROS-2 BCs (lower; filled dots: LMC; crosses: SMC) and SO/FO objects (upper; open circles: LMC; stars: SMC).}
\label{fig:Petersen2}
\end{figure}

Figure \ref{fig:Petersen2} displays the period ratio $P_{10} = P_{1}/P_{0}$ vs. ${\rm log} P_0$, a
Petersen diagram with all FO/F and SO/FO EROS-2 objects. While SO/FO Magellanic objects are mixed in the upper part of the diagram, a clear separation between FO/F LMC and SMC BCs is visible in the lower part. This was previously noted by \citet{1997A&A...321L...5B} who explained from a theoretical point of view that the observed period ratio results from a combination of two opposite effects: (i) the lower the metallicity, the smaller the opacity bump which affects the pulsation; (ii) at lower metallicity the structure of the star changes: at fixed mass, the luminosity increases and, therefore, the period ratio decreases. Moreover, the position of the blue loop of an evolutionary track of a Cepheid with respect to the instability strip is different in the LMC and in the SMC because of their different metallicity. As a consequence, lower pulsation periods are favoured in the SMC, while the highest periods are found in the LMC.

\begin{figure}
 \resizebox{\columnwidth}{!}{\includegraphics{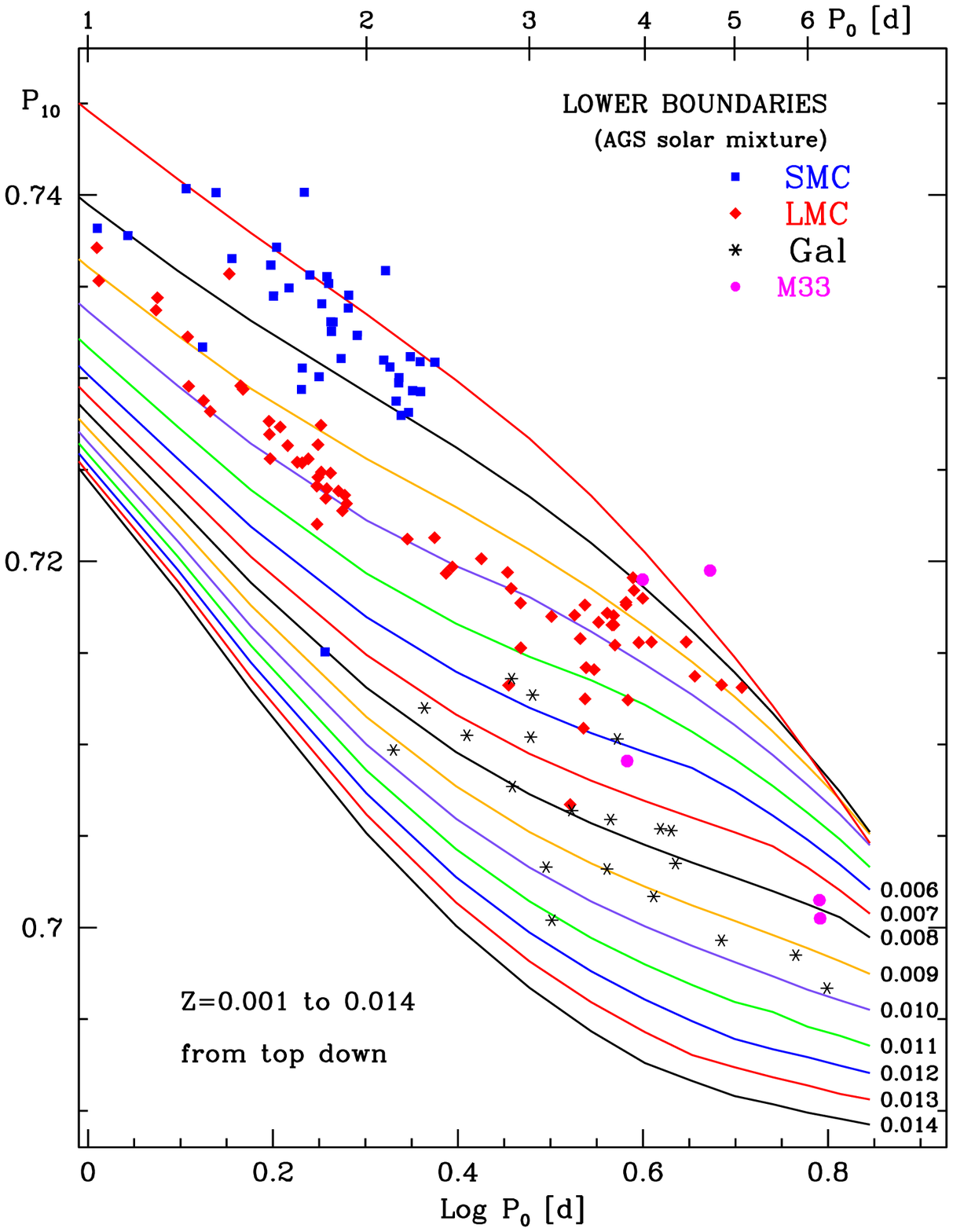}}
\caption{Composite Petersen diagram ($P_{10} = P_1/P_0$ vs. $P_0$) of the FO/F BCs from EROS-2 compared to other known stars. The lines delimit the periods for which both F and FO are linearly unstable at given $Z$ (lower boundary in $Z$ for given $P_{10}$ and $P_0$). The metallicity increases
 downward from $Z$ = 0.001 (top line) to 0.014 in steps of 0.001. Red rhombuses: 
EROS-2, LMC; blue squares: EROS-2, SMC; black stars: Galactic objects \citep{2004yCat.2250....0S}; pink circles: M33 objects \citep{2006ApJ...653L.101B}.}
\label{fig:Petersen1}
\end{figure}

\begin{figure}
 \resizebox{\columnwidth}{!}{\includegraphics{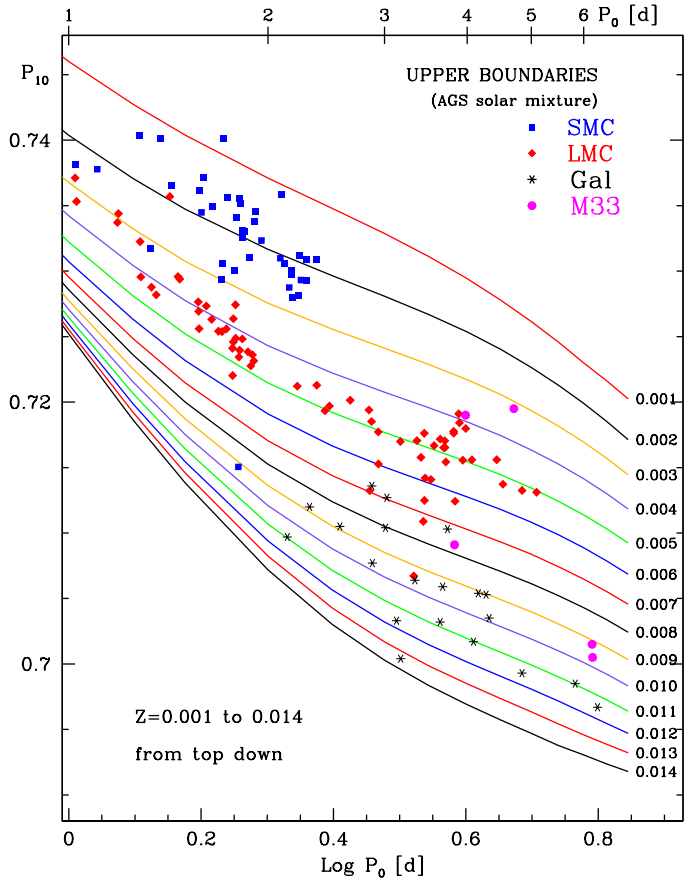}}
\caption{As Fig. \ref{fig:Petersen1} (same symbols), but showing the upper boundary in $Z$ for which models are unstable for given $P_{0}$ and $P_{10}$.}
\label{fig:Petersen1bis}
\end{figure}

The metallicity of FO/F BCs has been recently reexamined by
\citet{2007ApJ...660..723B}.  It has been known for some time that the location
of a BC in the Petersen diagram correlates strongly with metallicity $Z$.
We plotted all our LMC and SMC BCs on such a Petersen diagram
in Figs.~\ref{fig:Petersen1} and \ref{fig:Petersen1bis}.

\citet{2007ApJ...660..723B} developed a convenient tool for determining
metallicity limits, $Z_{\mathrm min}$ and $Z_{\mathrm max}$ from the 
location of a BC in the Petersen diagram, based only on accurate observed pulsation periods. From linear nonadiabatic convective stellar models they
constructed a composite Petersen diagram (Figs.~\ref{fig:Petersen1} and \ref{fig:Petersen1bis}), in which the isometallicity lines delimit the lower and upper boundaries, respectively, of the range in which both F
and FO modes are linearly unstable; this is the linear criterion for beat
pulsations.  

Thus, if a given BC falls below the upper period boundary for a certain 
value of $Z$ in Fig. \ref{fig:Petersen1}, and above the corresponding lower period boundary for the same $Z$ in Fig. \ref{fig:Petersen1bis},
then this is a possible value of $Z$ for this star. Thus, for example, for
the LMC BC with log $P_0 = 0.521$ and $P_{10} = 0.7067$,
Fig. \ref{fig:Petersen1} yields $Z_{\mathrm min} \sim 0.0077$, while 
Fig. \ref{fig:Petersen1bis} gives $Z_{\mathrm max}\sim 0.0093$. It is interesting to note that the observational data follow the theoretical isometallicity curves rather than a simple linear correlation.

\begin{figure}
  \resizebox{\hsize}{!}{\includegraphics{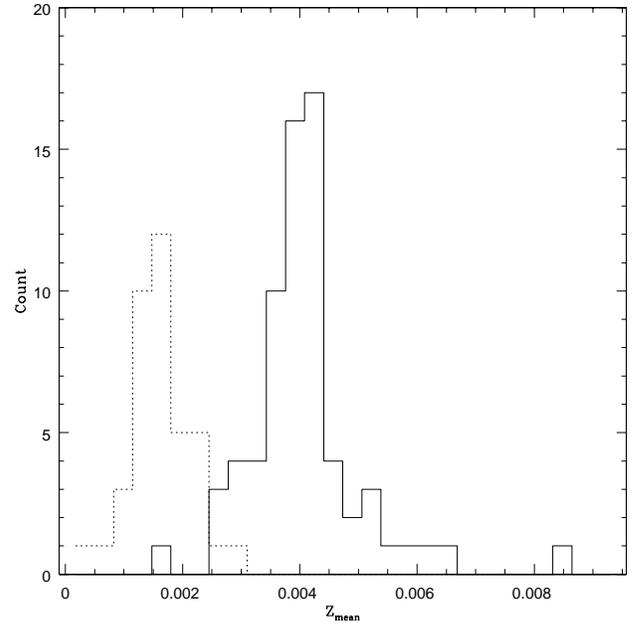}}
\caption{Distribution of mean metallicities of FO/F BCs in the Magellanic Clouds. Solid line: LMC; dashed line: SMC. The bin size is 3.25$\times 
10^{-4}$, corresponding to the mean error of $Z_{\mathrm mean}$.}
\label{fig:10842fg6} 
 \end{figure}

In practice, one interpolates among the isometallicity lines with a simple 
code.  Using this approach, we have computed the metallicities of our BCs,
and give the mean $Z_{\mathrm mean}$ together with $Z_{\mathrm min}$ and $Z_{\mathrm max}$ in the electronic tables.
Figure \ref{fig:10842fg6} shows the distribution of mean metallicity for the FO/F BCs in the Magellanic Clouds. The size of the bins is 3.25$\times 10^{-4}$, corresponding to the mean uncertainties in $Z_{\mathrm mean}$ which is $5.2 \times 10^{-4}$ in the LMC and $2.3 \times 10^{-4}$ in the SMC. 

A comparison of these data with other determinations is not straightforward, because previous spectroscopic measurements such as those by \citet{2005AJ....129.1465C} or \citet{2001A&A...373..159C} report [Fe/H] as obtained for stars from different populations, thus from different stages in the chemical evolution of the host galaxy than that of the young BCs. Also, we determine $Z$, not [Fe/H], and allowance for the non-Solar abundance ratios in the LMC and SMC \citep{2008A&A...480..379P} would be required to make a direct comparison. 

Three low-luminosity objects with periods $\sim 0.5$~d fall at the
left boundary of the Petersen diagrams: \object{lm0316m27403}, \object{lm0470n7559} and \object{lm0682m15849}
 (Fig. \ref{fig:Petersen2}).  They also appear to have 
$Z> 0.026$, which is not realistic.
This therefore seems to rule out that they are very low mass Cepheids in the
first crossing phase, as argued by \citet{1998ApJ...499L.205B}.  In fact, 
they may not be bona fide Cepheids within the range of our
calculated models.  However, we recall that the only assumption in our computations is that they are centrally
condensed, like Cepheids, so that they can be represented by envelope
models.  It could therefore be that these stars are less evolved.
Concerning the 4 BCs (2 LMC + 2 SMC) not represented on Figs. \ref{fig:Petersen1} and \ref{fig:Petersen1bis} as they are outside the range of $P_{10}$ on that figures, we did not find any particular behaviour. Therefore we suggest, that these BCs may be used to localized zone in the Magellanic Clouds of lower and higher metallicity than average.

\section{Discussion of the uncertainties in $Z$}
We comment here on the three main sources of uncertainty in the inferred metallicities. 

\subsection{The use of a linear Beat Cepheid criterion}

In Fig.~\ref{fig:lgP0_Z} we show the derived $Z$ values as a function of
log $P_0$.  The error bars reflect the consequences of using the linear
criterion for beat behavior, i.e. simultaneous instability in the fundamental
and first overtone modes.  A comparison of Figs.~\ref{fig:Petersen1}
and \ref{fig:Petersen1bis}, which show the upper and lower boundaries of $P_{10}$ as a function of $Z$, indicates that the uncertainties in $Z$ become very large for the longer-period BCs.  It is indeed well known that the linear BC criterion is far too relaxed, so that real BCs occupy only a
subregion of the corresponding parameter space \citep{1974ApJ...192..139S, 1986ApJ...308..661B}.

The obvious remedy is to perform nonlinear BC model calculations. This is 
a very computer intensive and time consuming endeavour, but is now under
completion (Szab{\'o} \& Buchler 2009, in preparation).  The work to date shows that the width of the double-mode instability
strip is narrower by about a factor of 5 in log $T$ in a theoretical HR
diagram.  In a diagram like Fig.~\ref{fig:Petersen1bis} this translates into a spread in $P_{10}$ of less than 0.0005; moreover, this is relatively uniform in log $P_0$, i.e. does not open up towards the long periods as 
with the linear criterion. It is also found that, overall, the nonlinear range is closer to the upper
linear boundary (Fig.~\ref{fig:Petersen1bis}) than the lower one (Fig.~\ref{fig:Petersen1}). On average, the nonlinear values of $Z$ are
therefore expected to be somewhat higher than the linear ones.

\subsection{The chemical composition}
Our models use OPAL opacities computed with the revised solar composition 
of \citet{2005ASPC..336...25A}.  This causes a substantial change in the inferred metallicity as compared to that obtained with the older \citet{1993oee..conf...14G} solar composition used
by \citet{2008ApJ...680.1412B}.  The inferred average ('linear') $Z$ in the LMC is reduced from 0.0070 to 0.0044 and that of the SMC from 0.0028 to
0.0018.  This is of course a gross overestimate of the true uncertainty in
metallicity, because any future revision of the composition is expected to be
smaller than the difference between \citet{2005ASPC..336...25A} and \citet{1993oee..conf...14G}.  We also note that Magellanic Cloud stars in 
general do not have Solar abundance ratios \citep{2008A&A...480..379P} 
as we have assumed in our models, and we do not know how much this might change the results.

\subsection{Modelling uncertainties}

\begin{figure*}
  \resizebox{\hsize}{!}{\includegraphics{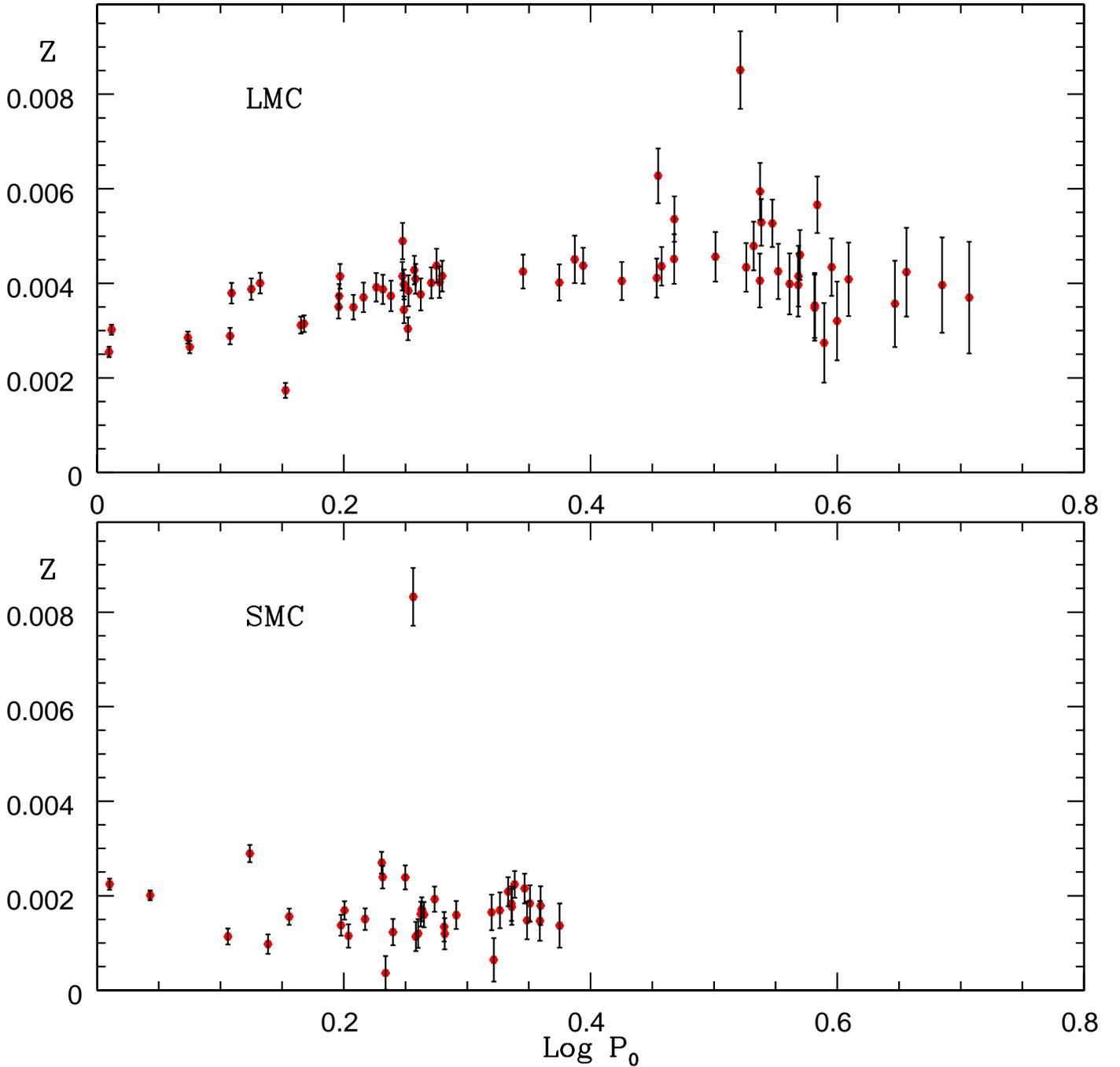}}
\caption{Plot of the metallicity $Z$ vs. the period of the fundamental mode. Error bars are derived from the lower and upper boundaries in Figs. \ref{fig:Petersen1} and \ref{fig:Petersen1bis}.}
\label{fig:lgP0_Z}
\end{figure*}

\citet{2008ApJ...680.1412B} has shown that the approximation of envelope models breaks down for the short-period Cepheids, and that it is necessary 
to compute full stellar models (which we are not equipped to do at this stage).  One may therefore wonder whether the lower metallicity at shorter periods seen in Fig.~\ref{fig:lgP0_Z} is a result of this uncertainty in the models, or whether there is a physical or evolutionary reason for the lower metallicity, e.g. \citet{2003A&A...409..491C}.

\section{Conclusion}
\label{sec:discuss}

We have reported a new survey of the BC populations in the Magellanic Clouds which complements the recent work of the OGLE group in the SMC \citep{2008arXiv0808.2210S} and extends it over a larger area in the LMC. Our new, large FO/F sample yields a new metallicity distribution for another stellar population in the LMC and the SMC. Together with previous studies of HII regions, planetary nebulae or red giants, this contributes to a more complete view of the metallicity of the Magellanic Clouds and its links to stellar evolution. We deduce from the present data a mean metallicty of 0.0045 in the LMC and 0.0018 in the SMC.

The demonstration that such BC stars are a useful and independent way to determine metallicities has considerable promise for future studies in the Local Group. In \object{M33}, 5 BCs were found among a sample of more than 2000 Cepheids \citep{2006ApJ...653L.101B}, and we have shown here that tens of BCs can be detected in the Magellanic Clouds, among a total population 
of 2000--3000 single-mode Cepheids (see \citet{2008arXiv0808.2210S} and Table \ref{tab:KS} above). The frequency of BCs in other galaxies of the Local Group is likely to be between these values. \object{M31} is the obvious first candidate for such a search.

 \begin{acknowledgements}
 JPB, JBM and ASC acknowledge financial support from the European Associated Laboratory "Astrophysics Poland-France". JRB and ASC gratefully acknowledge, respectively, the support of NSF and KBN (grants AST 07-07972 and OISE 04-17772 at UF, and 1P03D~02529 at UAM).
RSz acknowledges the support of a Hungarian E\"otv\"os Fellowship.
RSz and JRB are grateful to the Hungarian NIIF Supercomputing Facility
and to the UF High-Performance Computing Center for providing computational
resources and support.
 \end{acknowledgements}

\bibliographystyle{aa}
\bibliography{10842}

\end{document}